\documentclass[journal = jpcafh, manuscript=article]{achemso}
\usepackage{graphicx}
\usepackage{array, multirow}
\usepackage{amsmath}
\usepackage{multicol}
\usepackage{epstopdf}
\usepackage{color}
\usepackage{leftidx}
\SectionNumbersOn
\DeclareGraphicsExtensions{.eps}
\usepackage[caption=false]{subfig}
\usepackage{soul}
\newcommand{\subfigimg}[3][,]{%
  \setbox1=\hbox{\includegraphics[#1]{#3}}
  \leavevmode\rlap{\usebox1}
  \rlap{\hspace*{10pt}\raisebox{\dimexpr\ht1-2\baselineskip}{#2}}
  \phantom{\usebox1}
}
\newcommand{\onlinecite}[1]{\hspace{-1 ex} \nocite{#1}\citenum{#1}}

\title{Active Space Pair 2-Electron Reduced Density Matrix Theory for Strong Correlation}

\author{Kade Head-Marsden and David A. Mazziotti}
\email{damazz@uchicago.edu}

\affiliation{Department of Chemistry and The James Franck Institute, \\ The University of Chicago, Chicago, IL 60637 USA}

\date{Submitted March 4, 2020; Revised April 26, 2020}

\begin{document}

\begin{abstract}
 An active space variational calculation of the 2-electron reduced density matrix (2-RDM) is derived and implemented where the active orbitals are correlated within the pair approximation. The pair approximation considers only doubly occupied configurations of the wavefunction which enables the calculation of the 2-RDM at a computational cost of $\mathcal{O}(r^3)$. Calculations were performed both with the pair active space configuration interaction (PASCI) method and the pair active space self consistent field (PASSCF) method. The latter includes a mixing of the active and inactive orbitals through unitary transformations. The active-space pair 2-RDM method is applied to the nitrogen molecule, the p-benzyne diradical, a newly synthesized BisCobalt complex, and the nitrogenase cofactor FeMoco.  The FeMoco molecule is treated in a [120,120] active space.  Fractional occupations are recovered in each of these systems, indicating the detection and recovery of strong electron correlation.
\end{abstract}

\section{Introduction}

Recent work uses active space selection in conjunction with the variational 2-RDM method to capture strong electron correlation in a variety of molecules of chemical interest.\cite{Schlimgen2016, Montgomery2018, McIsaac2017} An active space is a set of orbitals within the molecule that are correlated. Increasing the size of the active space alters the amount of electron correlation that can be captured with larger active spaces producing more accurate results. Because the variational 2-RDM method with the 2-positive $N$-representability conditions\cite{Mazziotti2007a, Garrod1975, Erdahl1979, Mazziotti2001a, Nakata2001, Zhao2004, Mazziotti2002, Mazziotti2006, Gidofalvi2008, Pelzer2011, Verstichel2012, Fosso-Tande2016, Schlimgen2016, Mazziotti2016, Coleman1963, Erdahl1978, Kummer1967, Mazziotti2001a, Vandenberghe1996, Mazziotti2004, Mazziotti2007, Mazziotti2011, Piris2013} scales as $\mathcal{O}(r_{a}^6)$ where $r_{a}$ is the number of active orbitals, it can treat much larger active spaces than conventional configuration interaction, which scales exponentially with $r_{a}$.   \textcolor{black}{To further reduce the scaling of the variational 2-RDM method, h}ere we combine the recently developed pair variational 2-RDM method\cite{Poelmans2015, Naftchi-Ardebili2011, Head-Marsden2017, Alcoba2018, Alcoba2018_2, Alcoba2019} with active space methods, \textcolor{black}{the pair space equivalent to both complete active space configuration interaction (CASCI) and complete active space self consistent field (CASSCF)}, to generate an $O(r_{a}^{3})$ method that can efficiently treat strong correlation in molecules with \textcolor{black}{the potential to treat significantly larger active spaces than current methods.}

The \textcolor{black}{doubly-occupied configuration interaction (DOCI) or} pair space restricts the wavefunction to include only doubly-occupied determinants; however, in traditional wavefunction methods this approximation alone still scales exponentially with system size.\cite{Weinhold1967, Szabo1996} While the variational 2-RDM method reduces the computational scaling to polynomial, there are also several wavefunction approximations for decreasing the doubly occupied configuration interaction scaling to polynomial such as antisymmetric product of one-reference-orbital geminals (AP1roG) and pair coupled cluster doubles (pCCD).\cite{Limacher2013, Boguslawski2014, Boguslawski2014a, Tecmer2014, Boguslawski2015, Bytautas2011, Stein2014, Henderson2014, Henderson2015, Bulik2015, Shepherd2016} Moreover, some work has examined active space calculations in spaces of different seniority for small molecules.\cite{Byautas2011}

In this paper the pair variational 2-RDM theory with an active space self-consistent field (PASSCF) method is utilized where the active and inactive orbitals are iteratively rotated to decrease the energy. \textcolor{black}{Combining pair methods with active-inactive orbital rotations provides an efficient approach to treating orbital rotations which have limited many pair calculations to small molecular sizes.  The existing algorithms for self consistent field (SCF) in CASSCF can be utilized without modification for efficient orbital rotations.  Active-active rotations, while not explicitly treated by many SCF algorithms for CASSCF, are still indirectly included through two or more active-inactive orbital rotations.} We benchmark this method using \textcolor{black}{the dissociation of a nitrogen dimer, a p-benzyne diradical, a recently synthesized bis-cobalt complex and a recently studied iron complex, FeMoco.\cite{Montgomery2018}}

\section{Theory}

The energy of an $N$-electron system can be expressed as,\cite{Mazziotti2012, Mazziotti2007a, Coleman2000, Davidson1976, Valdemoro1992, Nakatsuji1996, Mazziotti1998, Lowdin1955, Mayer1955}
\begin{equation}
 E = \sum_{ijkl} {^2}K^{ij}_{kl}\ \leftidx{^2}D^{ij}_{kl},
\end{equation}
where ${^2}K^{ij}_{kl}$ is the two-electron Hamiltonian given by,
\begin{equation}
 ^2K_{kl}^{ij} = \frac{4}{N-1} {^1}K^i_k\wedge \delta^j_l + ^2V_{kl}^{ij},
\end{equation}
in which, ${^1}K^i_k$ and ${^2}V_{kl}^{ij}$ are one- and two-electron matrices containing the one- and two-electron integrals and $\wedge$ is the Grassmann wedge product.~\cite{Mazziotti1998, Slebodzinski1970}

\textcolor{black}{Semidefinite programming can be used to minimize the ground-state energy with respect to the 2-RDM subject to the following approximate $N$-representability constraints,\cite{Schlimgen2016, Mazziotti2016, Vandenberghe1996, Mazziotti2007a, Mazziotti2004, Mazziotti2007, Mazziotti2011} referred to as the DQG conditions,}\cite{Coleman1963, Mazziotti2012a, Garrod1964, Mazziotti2012, Mazziotti2007a, Coleman2000, Davidson1976, Valdemoro1992, Nakatsuji1996, Mazziotti1998, Lowdin1955, Mayer1955, Coleman1963, Garrod1964, Kummer1967, Erdahl1978, Mazziotti2001a, Mazziotti2012a, Garrod1964, Fukuda2007}
\begin{eqnarray}
 ^2D &\succeq& 0\\
 ^2Q &\succeq& 0\\
 ^2G &\succeq& 0,
\end{eqnarray}
where $^2D$, $^2Q$, and $^2G$ are the two-particle, two-hole, and particle-hole density matrices respectively whose matrix elements are defined by 
\begin{eqnarray}
^2D^{ij}_{kl} &= \langle \Psi \lvert \hat{a}_i^{\dagger} \hat{a}_j^{\dagger} \hat{a}_k \hat{a}_l \rvert \Psi \rangle\\
^{2} Q_{kl}^{ij}  &= \langle \Psi \vert \hat{a}_i\hat{a}_j\hat{a}_l^{\dagger}\hat{a}_k^{\dagger}\vert \Psi \rangle \\
^{2} G_{kl}^{ij}  &= \langle \Psi \vert \hat{a}_i^{\dagger}\hat{a}_j\hat{a}_l^{\dagger}\hat{a}_k\vert \Psi \rangle.
\end{eqnarray}
\textcolor{black}{The variational 2-RDM method with necessary $N$-representability conditions is a lower-bound method in that the computed energy is a lower bound to the ground-state energy in a given finite basis set.}

In the pair approximation, \textcolor{black}{these three matrices take a block-diagonal form.} For the 2-particle and 2-hole RDMs, these \textcolor{black}{structures} consist of one $r\times r$ block and $r$-choose-two 1$\times$1 blocks; for example, for $^{2} D$ they are given by: \cite{Head-Marsden2017,Coleman2000}
\begin{equation}
  \begin{pmatrix}
  \langle \hat{a}^{\dagger}_{i\alpha}\hat{a}^{\dagger}_{i\beta}\hat{a}_{i\beta}\hat{a}_{i\alpha} \rangle &  \hdots & \langle \hat{a}^{\dagger}_{i\alpha}\hat{a}^{\dagger}_{i\beta}\hat{a}_{k\beta}\hat{a}_{k\alpha} \rangle\\
    \vdots &  \ddots & \vdots \\
   \langle \hat{a}^{\dagger}_{k\alpha}\hat{a}^{\dagger}_{k\beta}\hat{a}_{i\beta}\hat{a}_{i\alpha} \rangle & \hdots & \langle \hat{a}^{\dagger}_{k\alpha}\hat{a}^{\dagger}_{k\beta}\hat{a}_{k\beta}\hat{a}_{k\alpha} \rangle\\
 \end{pmatrix},
\end{equation}
and
\begin{equation}
\begin{pmatrix}
 \langle \hat{a}^{\dagger}_{i\alpha}\hat{a}^{\dagger}_{j\beta}\hat{a}_{j\beta}\hat{a}_{i\alpha} \rangle
\end{pmatrix}.
\end{equation}
The \textcolor{black}{pair} structure of the particle-hole RDM is similar but slightly more complex in structure with one $r\times r$ block and $r$-choose-two 2$\times$2 blocks given by\cite{Head-Marsden2017}
\begin{equation}
 \begin{pmatrix}
   \langle \hat{a}^{\dagger}_{i\alpha}\hat{a}_{i\alpha}\hat{a}^{\dagger}_{i\alpha}\hat{a}_{i\alpha} \rangle & \hdots & \langle \hat{a}^{\dagger}_{i\alpha}\hat{a}_{i\alpha}\hat{a}^{\dagger}_{k\alpha}\hat{a}_{k\alpha} \rangle\\
   \vdots & \ddots & \vdots & \\
   \langle \hat{a}^{\dagger}_{k\alpha}\hat{a}_{k\alpha}\hat{a}^{\dagger}_{i\alpha}\hat{a}_{i\alpha} \rangle & \hdots  & \langle \hat{a}^{\dagger}_{k\alpha}\hat{a}_{k\alpha}\hat{a}^{\dagger}_{k\alpha}\hat{a}_{k\alpha} \rangle\\
 \end{pmatrix},
\end{equation}
and
\begin{equation}
\begin{pmatrix}
 \langle \hat{a}^{\dagger}_{i\alpha}\hat{a}_{j\beta}\hat{a}^{\dagger}_{j\beta}\hat{a}_{i\alpha} \rangle & \langle \hat{a}^{\dagger}_{i\alpha}\hat{a}_{j\beta}\hat{a}^{\dagger}_{i\beta}\hat{a}_{j\alpha} \rangle \\
  \langle \hat{a}^{\dagger}_{j\alpha}\hat{a}_{i\beta}\hat{a}^{\dagger}_{j\beta}\hat{a}_{i\alpha} \rangle & \langle \hat{a}^{\dagger}_{j\alpha}\hat{a}_{i\beta}\hat{a}^{\dagger}_{i\beta}\hat{a}_{j\alpha} \rangle
\end{pmatrix}.
\end{equation}
With the block diagonal forms of ${^2}D$, ${^2Q}$, and ${^2}G$, the scaling of the \textcolor{black}{pair} variational 2-RDM method is $\mathcal{O}(r^3)$.

Here, we use the pair theory within the context of an active space calculation. Performing a calculation in an active space consists of choosing $N$ electrons in $r$ orbitals to correlate while treating the remainder of the electrons and orbitals at a mean-field level of theory.\cite{Roos1987} \textcolor{black}{Importantly, in the case of an active-space calculation the lower bound to the energy is with respect to the configuration interaction in the active space.} We consider two primary active space methods within the \textcolor{black}{pair} variational 2-RDM framework. First, we consider a method \textcolor{black}{similar} to complete active space configuration interaction (CASCI), \textcolor{black}{in that} the active space is computed with respect to the Hartree-Fock canonical molecular orbitals. Second, we explore a method equivalent to the complete active space self consistent field (CASSCF) method,\cite{Werner1985} where active and inactive orbitals are iteratively rotated through a self-consistent field method. We use the second-order orbital optimization method described in Ref.~\onlinecite{Sun2017}.

\textcolor{black}{The advantage of combining pair methods with active-inactive orbital rotations is that all of the existing methods for efficient rotations~\cite{Sun2017} can be utilized without modification.  Results indicate that this combination avoids some of the orbital optimization issues previously reported for pair theories.  Furthermore, the active-inactive rotations through two or more rotations between the active and inactive orbitals indirectly include mixing of active orbitals.} The computational cost of the \textcolor{black}{PASCI} calculation remains $\mathcal{O}(r^3)$, while the orbital rotations in the \textcolor{black}{PASSCF} calculations increases the scaling to $\mathcal{O}(r^5)$.

\section{Results}

In this section we will discuss our computational methodology followed by applications to the dissociation of a nitrogen dimer, a p-benzyne diradical, a newly synthesized Bis-Cobalt complex and FeMoco in Secs. \ref{subsec:compmeth},\ref{subsec:N2}, \ref{subsec:pben}, \ref{subsec:biscobalt}, and \ref{subsec:femoco} respectively.

\subsection{Computational Methodology}
\label{subsec:compmeth}
We have implemented the active-space pair variational 2-RDM method in the Maple Quantum Chemistry Package, an add-on package for electronic structure in the computer algebra system Maple.\cite{RDMChem, Maple2019}  All calculations employ the 2-positivity (or D, Q, and G) $N$-representability conditions.  The cc-pVDZ basis set is used for nitrogen and the p-benzyne diradical calculations with [10,8] and [6,6] active spaces respectively.\cite{Dunning1989}  The 6-31g basis set is used for the Bis-Cobalt complex calculations in a [12,10] active space.\cite{Hehre1972} Finally, both STO-3G and cc-pVDZ basis sets are employed for the FeMoco calculations in [30,30], [60,60], [90,90], and [120,120] active spaces.\cite{Hehre1969,Pietro1983,Dobbs1987}\textcolor{black}{All calculations are performed without symmetry.} \textcolor{black}{The occupations are the eigenvalues of the 1-RDM, the natural-orbital occupations.}

\subsection{Nitrogen Dissociation}
\label{subsec:N2}
\begin{figure}[h!]
   \includegraphics[width = 0.9\textwidth]{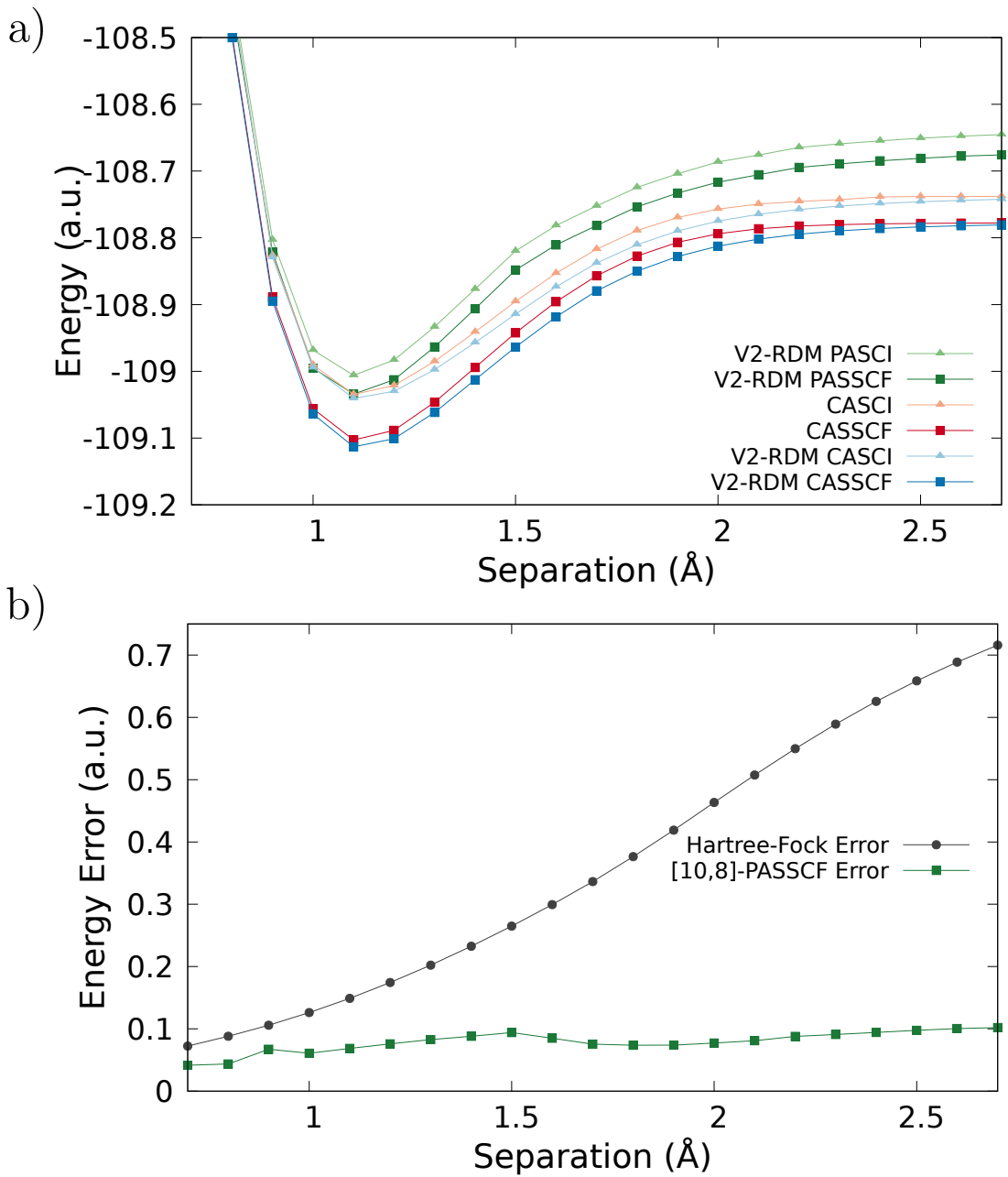}
  \caption{a) The dissociation of ${\rm N}_{2}$ using variational 2-RDM \textcolor{black}{PASCI} (light green triangles), variational 2-RDM \textcolor{black}{PASSCF} (green squares), CASCI (orange triangles), CASSCF (red squares), variational 2-RDM CASCI (light blue triangles), and variational 2-RDM CASSCF (blue squares). b) The energy error taken as the absolute value of the deviation from the CASSCF energy for  Hartree-Fock (grey) and variational 2-RDM PASSCF (light green) for various bond lengths of the N$_2$ molecule.}
  \label{N_diss}
\end{figure}

The nitrogen dimer is a known example of fractional occupations as it dissociates. In Fig.~\ref{N_diss} a) we compare the variational 2-RDM CASCI and CASSCF methods, the traditional wave function CASCI and CASSCF methods, and the 2-RDM PASCI and PASSCF methods. The error for the Hartree-Fock method as well as for the variational 2-RDM PASSCF method relative to the CASSCF energy is shown in Fig.~\ref{N_diss} b). For the same methods, the occupation numbers of the N$_2$ dimer at 1.2~\AA\ and 2.0~\AA\ are shown in Table ~\ref{N2_diss_tab}.

From the data in Figure~\ref{N_diss}, the approximate dissociation energy (well depth) is calculated by taking the absolute value of the difference between the energy at 2.7~\AA and the equilibrium 1.1~\AA. The CASSCF, 2-RDM CASSCF, and 2-RDM PASSCF produce dissociation energies of 325~mHartrees, 332~mHartrees, and 358~mHartrees respectively. As a point of comparison, the Hartree-Fock well depth, which is not shown in Figure~\ref{N_diss} is 892~mHartrees. This demonstrates the pair theory's ability to produce realistic potential energy surfaces, especially in the regions of multireference correlation.

\begin{table}[h!]
  \caption{The occupation numbers of N$_2$ at 1.2~\AA\ and 2.0~\AA\ separation using CASCI and CASSCF, variational 2-RDM CASCI and CASSCF, and variational 2-RDM \textcolor{black}{PASCI and PASSCF.}}
  \begin{tabular}{c  c c c c c c}
  Sep.  & \multicolumn{2}{c}{Wave Function} &  \multicolumn{4}{c}{Variational 2-RDM}\\
  \cline{2-3} \cline{4-7}
   (\AA) & CASCI & CASSCF & CASCI & CASSCF & \textcolor{black}{PASCI} & \textcolor{black}{PASSCF}\\
   \hline
1.2 & 1.995 & 1.995 & 1.994 & 1.985 & 2.000 & 1.998\\
 & 1.993 & 1.989 & 1.983 & 1.984 & 1.989 & 1.989\\
 & 1.988 & 1.974 & 1.988 & 1.976 & 1.997 & 1.993\\
 & 1.927 & 1.922 & 1.918 & 1.910 & 1.936 & 1.920\\
 & 1.927 & 1.922 & 1.918 & 1.910 & 1.936 & 1.920\\
 & 0.081 & 0.086 & 0.094 & 0.102 & 0.070 & 0.087\\
 & 0.081 & 0.086 & 0.094 & 0.102 & 0.070 & 0.087\\
 & 0.008 & 0.028 & 0.011 & 0.032 & 0.002 & 0.008\\
2.0 & 1.998 & 1.999 & 1.989 & 1.988 & 1.999 & 1.999\\
 & 1.995 & 1.995 & 1.985 & 1.986 & 1.997 & 1.997\\
 & 1.674 & 1.659 & 1.714 & 1.704 & 1.780 & 1.783\\
 & 1.330 & 1.316 & 1.357 & 1.348 & 1.347 & 1.334\\
 & 1.330 & 1.316 & 1.357 & 1.348 & 1.347 & 1.334\\
 & 0.671 & 0.686 & 0.651 & 0.661 & 0.655 & 0.667\\
 & 0.671 & 0.686 & 0.651 & 0.661 & 0.655 & 0.667\\
 & 0.329 & 0.345 & 0.295 & 0.306 & 0.222 & 0.218\\
  \end{tabular}
  \label{N2_diss_tab}
\end{table}

While Figure~\ref{N_diss} a) shows that the energy recovered by the pair methods is less than that recovered by the 2-RDM or wave function CAS methods, the energy errors in Figure~\ref{N_diss} b) show that pair methods still recover a significant portion of the correlation energy.  Orbital rotations in the \textcolor{black}{PASSCF} slightly decrease the energy as compared to the \textcolor{black}{PASCI}. As the two nitrogen atoms are separated, Table~\ref{N2_diss_tab} shows an increase in partial occupations for all methods, demonstrating the pair theory's ability to capture strong correlation and produce accurate dissociation curves.

\subsection{p-Benzyne Diradical}
\label{subsec:pben}

\begin{figure}[h!]
  \subfigimg[width = 0.45\textwidth, trim = 0 15cm 8.5cm 0, clip]{a)}{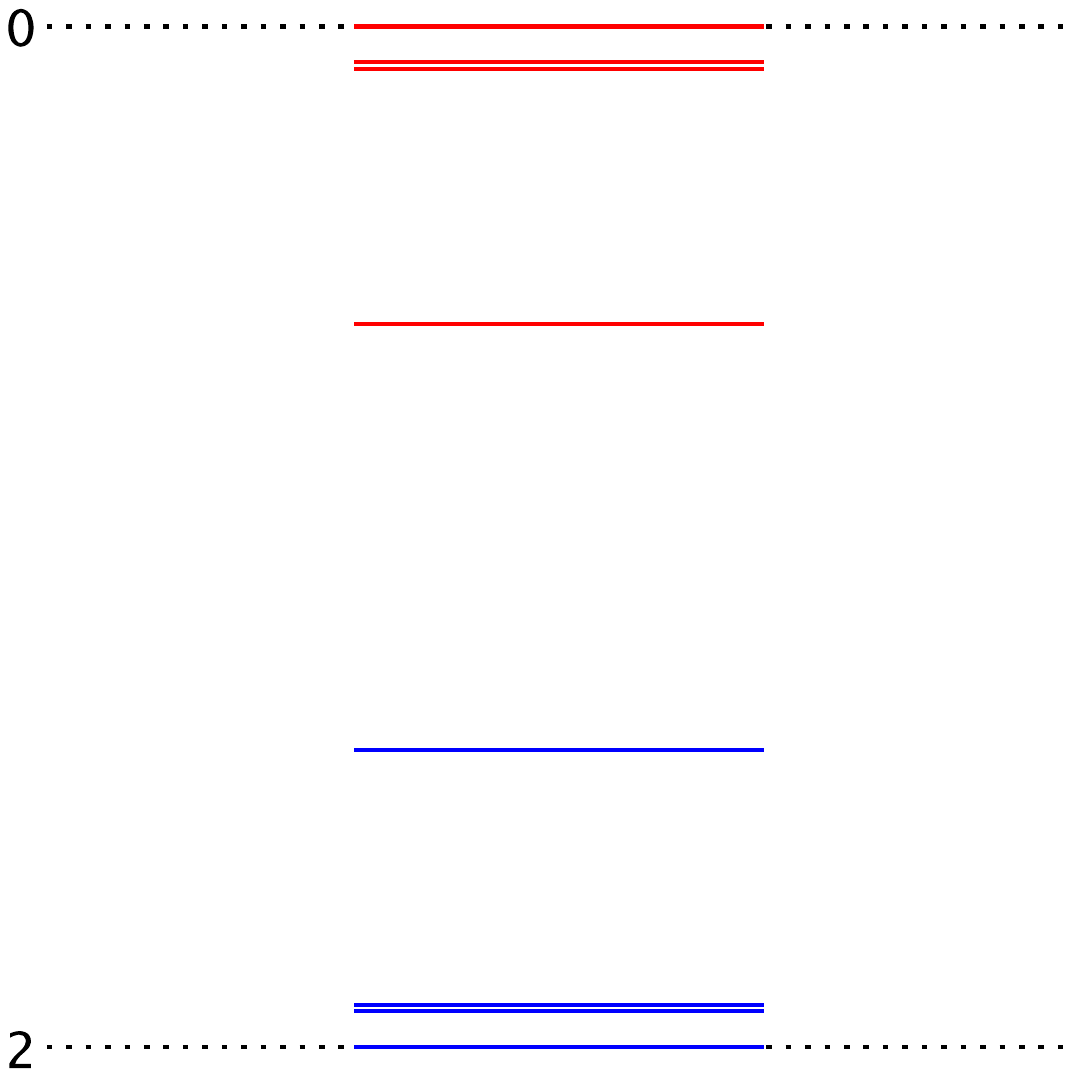}
  \subfigimg[width = 0.45\textwidth, trim = 0 15cm 8.5cm 0, clip]{b)}{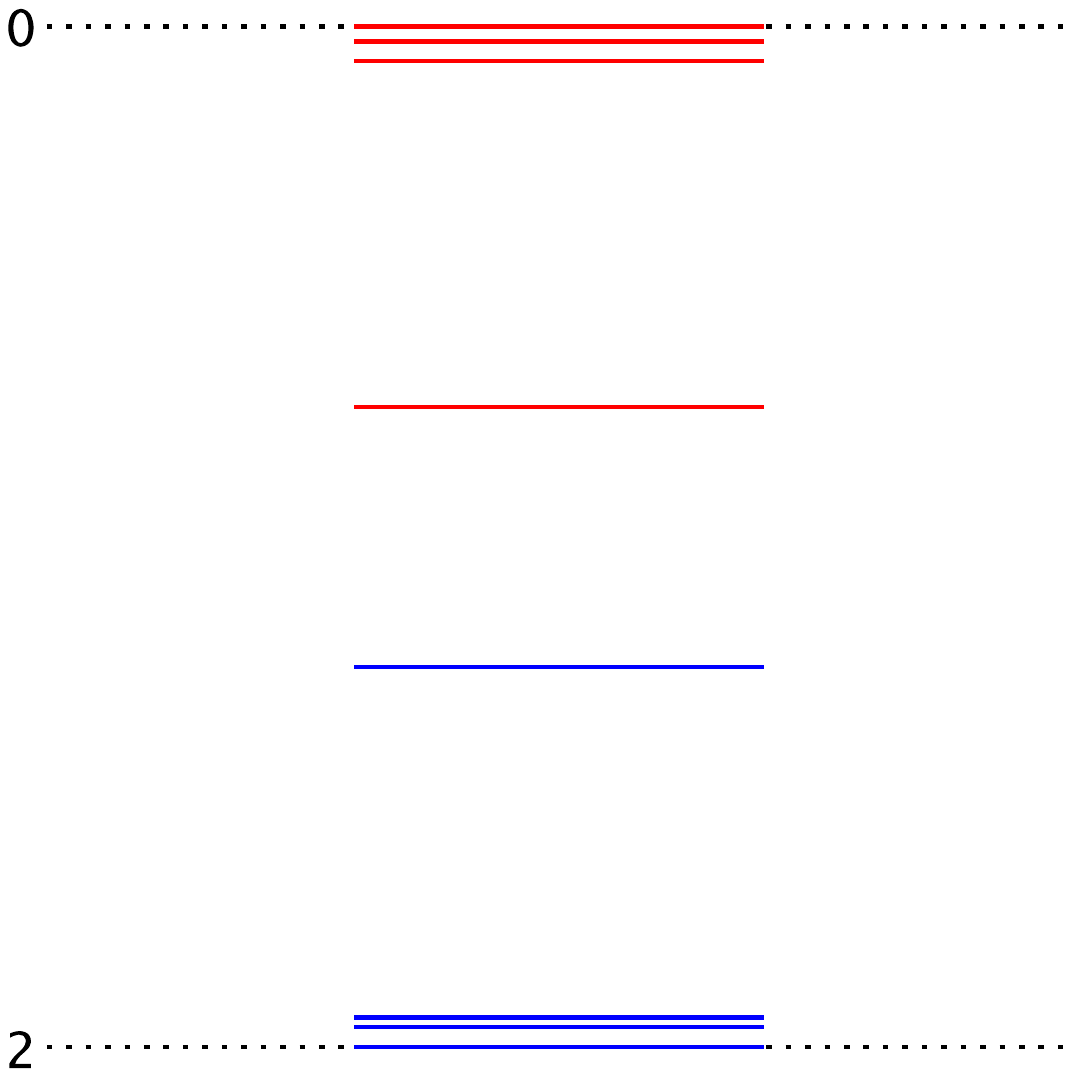}
  \caption{Molecular orbital occupations line plot for the p-benzyne diradical
  using the variational 2-RDM a) CASSCF and b) \textcolor{black}{PASSCF} methods.}
  \label{p-benzyne-lineplot}
\end{figure}

The Hartree-Fock, CASSCF, and both the variational 2-RDM CASSCF and PASSCF methods are used to calculate the lowest singlet and triplet energies for the p-benzyne diradical in the cc-pVDZ basis set as shown in Table~\ref{p-benzyne-energies}.  Even though the singlet is lower then the triplet experimentally by 3.8$\pm$0.3~kcal/mol,~\cite{Wenthold1998, Shee2019} the restricted Hartree-Fock calculation predicts the triplet to be lower than the singlet by 69.0~kcal/mol, showing that at least with correct spin symmetry all of the lowering of the singlet energy relative to the triplet energy is attributable to electron correlation.  The singlet-triplet gaps from HF, 2-RDM PASSCF, 2-RDM CASSCF (with DQG conditions), and CASSCF are -69.03, 1.26, 5.02, and 3.77~kcal/mol.  These results show that the pair approximation in PASSCF captures most of the  singlet state's correlation energy that lowers its total energy below that of the triplet state.  The 2-RDM PASSCF singlet-triplet gap is too small because the pair approximation does not capture all of the electron correlation while the 2-RDM CASSCF gap is slightly too large because the approximate $N$-representability conditions allow the singlet biradical to overcorrelate relative to the less correlated triplet state.  The occupation numbers of the singlet state using the variational 2-RDM PASSCF and \textcolor{black}{CASSCF  methods, presented in Fig.~\ref{p-benzyne-lineplot}}, show that both methods capture the state's biradical character, indicating that PAS is sufficient to capture the biradical character.


\begin{table}[h!]
  \caption{The energies of p-benzyne using Hartree-Fock, CASSCF, and both the variational 2-RDM CASSCF and PASSCF methods.}
  \begin{tabular}{c c  c  c c }
   & & & \multicolumn{2}{c}{Variational 2-RDM}\\
   \cline{2-5}
    & HF & CASSCF & CASSCF & PASSCF\\
   \hline
Singlet Energy (Hartrees) &-229.27 & -229.43 & -229.44 & -229.41\\
Triplet Energy (Hartrees) & -229.38 & -229.42 & -229.43 & -229.41\\
Singlet-Triplet Gap (kcal/mol) & -69.03 & 3.77 & 5.02 & 1.26\\
   \end{tabular}
  \label{p-benzyne-energies}
\end{table}


\subsection{Cobalt Complex}
\label{subsec:biscobalt}

\begin{figure}[h!]
  \includegraphics[width = 0.85\textwidth, trim = 0 0 0.2cm 0, clip]{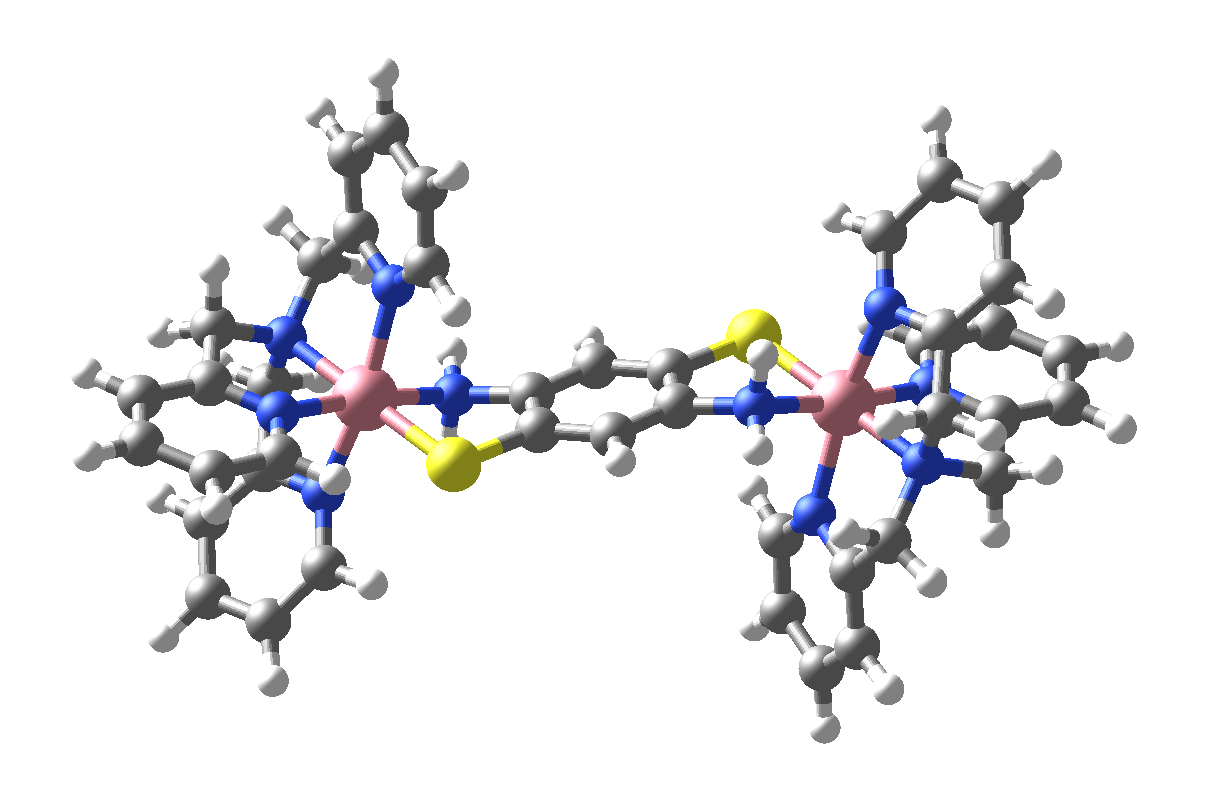}
  \caption{Bis-Cobalt complex where Carbon atoms are shown in grey, Hydrogen in white, Cobalt in pink, Sulfur in yellow, and Nitrogen in blue.\cite{RDMChem}}
  \label{Co2}
\end{figure}

Next, we consider a Bis-Cobalt complex [(CoTPA)$_2$DADT]$^{2+}$  (TPA is tris(2-pyridylmethyl)amine and DADT is 2,5-diaminobenzene-1,4-bis(thiolate)), shown in Fig.~\ref{Co2}, that has recently been synthesized and studied.\cite{Xie2019} Recent work has considered the effects of different linker molecules between the cobalt centers for tuning the amount of electron correlation for a variety of potential applications.\cite{Xie2019}

\begin{figure}[h!]
   \subfigimg[width = 0.5\textwidth]{a)}{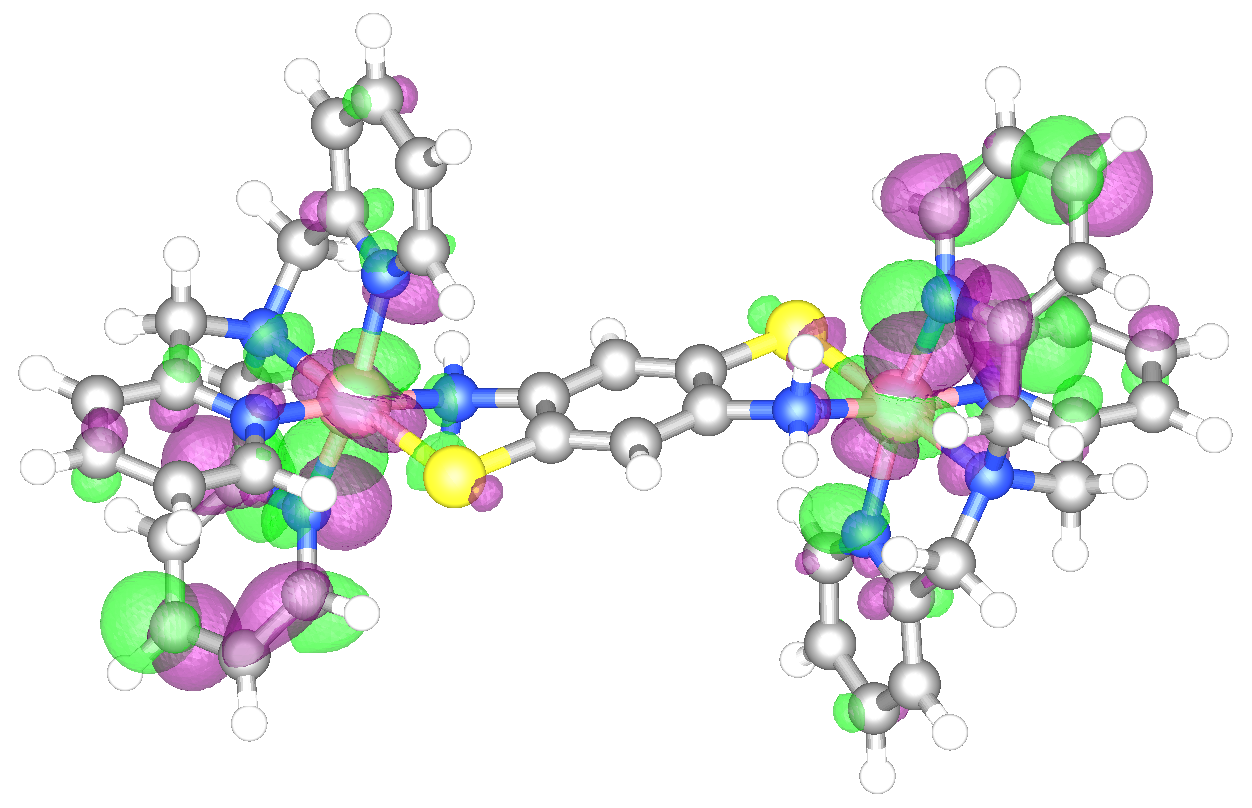}
   \subfigimg[width = 0.5\textwidth]{b)}{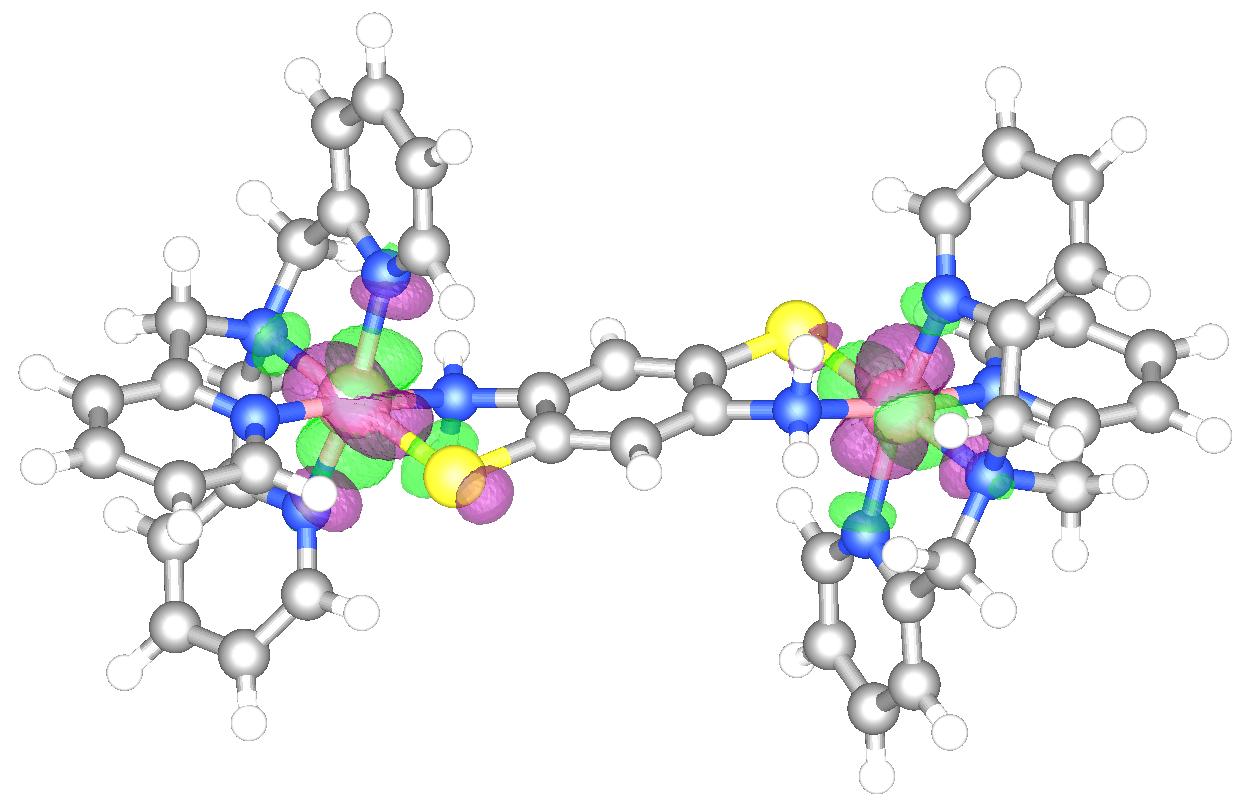}
  \caption{\textcolor{black}{Bis-cobalt complex $N/2$ (224) natural-orbital density with phases indicated by green and purple using a) Hartree-Fock and b) 2-RDM PASSCF.}}
  \label{Co2MOs}
\end{figure}

The \textcolor{black}{$N/2$ (224) natural orbital from Hartree-Fock and variational 2-RDM PASSCF are shown in Fig.~\ref{Co2MOs}. The $N/2$ natural orbital from Hartree-Fock shows significant delocalization of the orbital density over the ligands while the $N/2$ orbital from variational 2-RDM PASSCF reveals highly localized electron density on the two metal centers, which is consistent with the entanglement of an electron on each cobalt atom into a biradical.  The occupation numbers from variational 2-RDM PASCI and PASSCF show that $N/2$ and $N/2+1$ natural orbitals are half-filled, confirming the existence of a biradical.} \textcolor{black}{The correlation energy recovered is 208~mHartrees and 384~mHartrees for the 2-RDM \textcolor{black}{PASCI and PASSCF} calculations respectively.  The energy recovered by 2-RDM PASSCF is less than that recovered by 2-RDM CASSCF by only 6~mHartrees, revealing in this case that the pair approximation with SCF rotations recovers most of the electron correlation.}  The biradical character of the Bis-Cobalt complex is important to its magnetic properties, especially upon assembly into a larger crystalline solid.

\begin{table}[h!]
  \caption{The occupation numbers of Bis-Cobalt complex using variational 2-RDM PASCI and PASSCF with \textcolor{black}{12 electrons in 10 orbitals.} }
  \begin{tabular}{c c  c }
  MO Index & PASCI & PASSCF \\
   \hline
223 & 2.000 & 2.000\\
224 & 1.063 & 1.020\\
225 & 0.937 & 0.980\\
226 & 0.000 & 0.000\\

\end{tabular}
  \label{table:Co2Occ}
\end{table}

\subsection{FeMoco}
\label{subsec:femoco}

\begin{figure}[h!]
  \includegraphics[width = 0.45\textwidth]{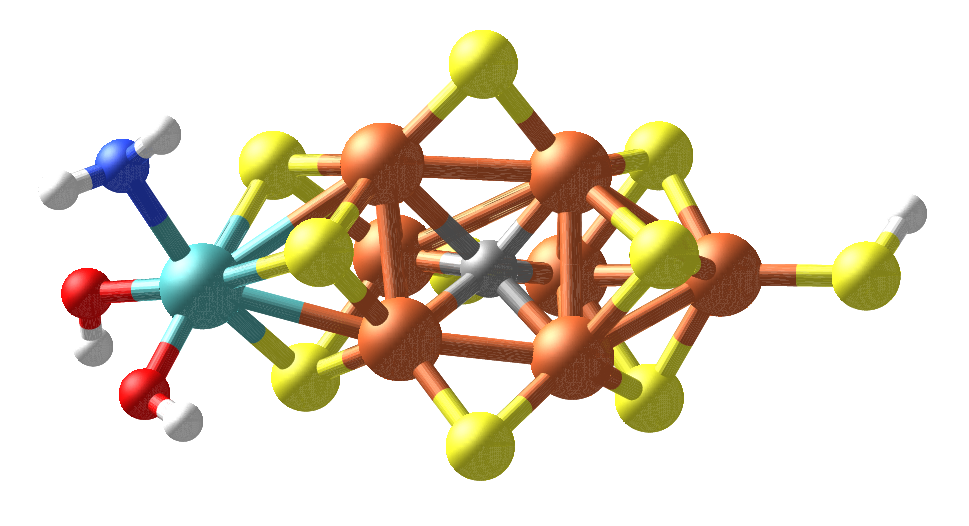}
  \caption{Modified FeMoco molecule where Molybdenum is shown in cyan, Sulfur in yellow, Iron in brown, Oxygen in red, Nitrogen in blue, Carbon in grey, and Hydrogen in white.\cite{RDMChem}}
  \label{femoco}
\end{figure}

Finally, we consider the modified FeMoco molecule, where the base chemical formula is MoFe$_7$S$_9$C, as shown in Fig.~\ref{femoco}.\cite{Montgomery2018} FeMoco is the active catalytic site in the reduction of nitrogen gas to ammonia during the process of nitrogen fixation.\cite{Burgess1996, Hoffman2014, Montgomery2018}

\textcolor{black}{The total energies and correlation energies for FeMoco, presented in Table~~\ref{table:FMnrgs}, were calculated in the STO-3G and cc-pVDZ basis sets in [30,30], [60,60], [90,90], and [120,120] active spaces.  Calculations were performed sequentially with the optimized orbitals from one active-space calculation being used to seed the orbitals of the next larger calculation.  We observe that strong electron correlation is recovered in the STO-3G basis set for all active space size while it is not recovered in the cc-pVDZ basis set even for the largest active space.  While the pair approximation is sufficient to capture some of the strong correlation in FeMoco in the STO-3G basis set, it is not sufficient to capture such correlation in the larger cc-pVDZ basis set.  These results are consistent with previous data on FeMoco from the variational 2-RDM method without the pair approximation.  In these earlier calculations it was seen that in larger basis sets larger active spaces on the order of [20,20] are required to observe the cross-over from a nearly Hartree-Fock solution to a highly correlated solution.  Here we see that it is not only the sizes of the basis set and the active space but also the degree of correlation supported by the electronic structure method that have a role in the competition between near-Hartree-Fock and strongly correlated solutions.  In the larger basis set the orbitals of the Hartree-Fock solution have many degrees of freedom that lower its energy to an extent that is difficult for the correlated solution to surpass without the full flexibility of a complete, non-pairing solution of the Schr{\"o}dinger equation.  The application of the pair approximation to FeMoco in these two basis sets is highly instructive because it reveals the subtle but important interplay of the basis set, active space, and correlation method in treating strong electron correlation.  Obtaining an accurate, correlated description of FeMoCO is important because the correlation affects properties from the atomic charges to the excited-state splittings that can influence its catalytic activity in nitrogen fixation.}

\begin{table}[h!]
  \caption{The total and correlation energies of FeMOCO in Hartrees using variational 2-RDM PASSCF using STO-3G and cc-pVDZ basis sets in [30,30], [60,60], [90,90], and [120,120] active spaces. }
  \begin{tabular}{c c  c c c }
  Active Space & \multicolumn{2}{c}{STO-3G} & \multicolumn{2}{c}{cc-pVDZ}\\
  \cline{2-3} \cline{4-5}
	&Total Energy	& Correlation Energy	& Total Energy	& Correlation Energy\\
	\hline
[30,30]	  & -16851.44 & -0.45 & -17030.92 & -0.20\\
{[}60,60] & -16851.52 & -0.53 & -17030.99 & -0.29\\
{[}90,90] & -16851.54 &	-0.55 & -17031.06 & -0.34\\
{[}120,120] &	-16851.56 & -0.57 & -17031.12 &	-0.40\\
\end{tabular}
  \label{table:FMnrgs}
\end{table}

\section{Discussion and Conclusion}

 Active space variational calculations of the 2-RDM are performed where the active orbitals are correlated within the pair approximation. The pair approximation, which consists of only considering $r/2$ pairs of orbitals in the wavefunction, greatly simplifies the structure of the 2-RDM. By invoking this approximation, the computational cost of the variational calculation of the 2-RDM constrained to the 2-positive (DQG) approximate $N$-representability conditions is reduced to $\mathcal{O}(r^3)$. Both \textcolor{black}{PASCI and PASSCF} calculations are considered in the treatment of N$_2$, a p-benzyne diradical, a Bis-Cobalt complex, and the nitrogenase cofactor, FeMoco. In each of these four systems, fractional occupation is observed, indicating the detection of strong electronic correlation. \textcolor{black}{The pair 2-RDM theory captures a certain family of electron correlation contained within the pair approximation, which in wave function terminology is the $N$-electron Hilbert space of all doubly occupied determinants (seniority zero).\cite{Poelmans2015, Naftchi-Ardebili2011, Head-Marsden2017, Alcoba2018, Alcoba2018_2, Alcoba2019, Limacher2013, Boguslawski2014, Boguslawski2014a, Tecmer2014, Boguslawski2015, Bytautas2011, Stein2014, Henderson2014, Henderson2015, Bulik2015, Shepherd2016, Byautas2011}  In some molecular systems the type of strong electron correlation may not be treatable within the pair approximation, such as the polyradical character in acene chains, shown in previous work~\cite{Head-Marsden2017}, and the calculation of FeMoco in the cc-pVDZ basis set, shown here.  These results are important because they not only show the limitations of pair theories in a more dramatic fashion than seen previously but also reveal the exquisite manner in which basis-set and active-spaces sizes as well as the nature of the correlation collectively play a role in determining whether the lowest-in-energy ground-state solution is nearly mean field or strongly correlated.  Despite the limitations in acene chains and FeMoco the present calculations still show that an active-space pair approximation with a self-consistent-field treatment of the inactive orbitals can capture strong electron correlation in a range of chemically relevant systems and that pair correlations have an important role in such systems.}  Due to the reduced computational cost and ability to capture strong correlation, the active-space pair 2-RDM methods provide a promising approach to treating molecular systems with large-scale active spaces beyond the Hartree-Fock limit.

\section{Acknowledgments}
D.A.M.   gratefully   acknowledges   the   U.S.   National Science  Foundation  Grant No.  CHE-1565638 and the  U.S.  Army Research Office (ARO) Grant No. W911NF-16-1-0152.

\bibliography{2019_Pair}

\providecommand{\latin}[1]{#1}
\makeatletter
\providecommand{\doi}
  {\begingroup\let\do\@makeother\dospecials
  \catcode`\{=1 \catcode`\}=2 \doi@aux}
\providecommand{\doi@aux}[1]{\endgroup\texttt{#1}}
\makeatother
\providecommand*\mcitethebibliography{\thebibliography}
\csname @ifundefined\endcsname{endmcitethebibliography}
  {\let\endmcitethebibliography\endthebibliography}{}
\begin{mcitethebibliography}{72}
\providecommand*\natexlab[1]{#1}
\providecommand*\mciteSetBstSublistMode[1]{}
\providecommand*\mciteSetBstMaxWidthForm[2]{}
\providecommand*\mciteBstWouldAddEndPuncttrue
  {\def\EndOfBibitem{\unskip.}}
\providecommand*\mciteBstWouldAddEndPunctfalse
  {\let\EndOfBibitem\relax}
\providecommand*\mciteSetBstMidEndSepPunct[3]{}
\providecommand*\mciteSetBstSublistLabelBeginEnd[3]{}
\providecommand*\EndOfBibitem{}
\mciteSetBstSublistMode{f}
\mciteSetBstMaxWidthForm{subitem}{(\alph{mcitesubitemcount})}
\mciteSetBstSublistLabelBeginEnd
  {\mcitemaxwidthsubitemform\space}
  {\relax}
  {\relax}

\bibitem[Schlimgen \latin{et~al.}(2016)Schlimgen, Heaps, and
  Mazziotti]{Schlimgen2016}
Schlimgen,~A.~W.; Heaps,~C.~W.; Mazziotti,~D.~A. Entangled Electrons Foil
  Synthesis of Elusive Low-Valent Vanadium Oxo Complex. \emph{J. Phys. Chem.
  Lett.} \textbf{2016}, \emph{7}, 627--631\relax
\mciteBstWouldAddEndPuncttrue
\mciteSetBstMidEndSepPunct{\mcitedefaultmidpunct}
{\mcitedefaultendpunct}{\mcitedefaultseppunct}\relax
\EndOfBibitem
\bibitem[Montgomery and Mazziotti(2018)Montgomery, and
  Mazziotti]{Montgomery2018}
Montgomery,~J.~M.; Mazziotti,~D.~A. Strong Electron Correlation in Nitrogenase
  Cofactor, FeMoco. \emph{J. Phys. Chem. A} \textbf{2018}, \emph{122},
  4988--4996\relax
\mciteBstWouldAddEndPuncttrue
\mciteSetBstMidEndSepPunct{\mcitedefaultmidpunct}
{\mcitedefaultendpunct}{\mcitedefaultseppunct}\relax
\EndOfBibitem
\bibitem[McIsaac and Mazziotti(2017)McIsaac, and Mazziotti]{McIsaac2017}
McIsaac,~A.~R.; Mazziotti,~D.~A. Ligand Non-innocence and Strong Correlation in
  Manganese Superoxide Dismutase Mimics. \emph{Physical Chemistry Chemical
  Physics} \textbf{2017}, \emph{19}, 4656--4660\relax
\mciteBstWouldAddEndPuncttrue
\mciteSetBstMidEndSepPunct{\mcitedefaultmidpunct}
{\mcitedefaultendpunct}{\mcitedefaultseppunct}\relax
\EndOfBibitem
\bibitem[Mazziotti(2007)]{Mazziotti2007a}
Mazziotti,~D.~A., Ed. \emph{Reduced-Density-Matrix Mechanics: With Application
  to Many-Electron Atoms and Molecules}; Advances in Chemical Physics 134;
  Wiley: New York, 2007\relax
\mciteBstWouldAddEndPuncttrue
\mciteSetBstMidEndSepPunct{\mcitedefaultmidpunct}
{\mcitedefaultendpunct}{\mcitedefaultseppunct}\relax
\EndOfBibitem
\bibitem[Garrod \latin{et~al.}(1975)Garrod, Mihailovic, and Rosina]{Garrod1975}
Garrod,~C.; Mihailovic,~M.; Rosina,~M. Variational Approach to 2-Body Density
  Matrix. \emph{J. Math. Phys.} \textbf{1975}, \emph{16}, 868--874\relax
\mciteBstWouldAddEndPuncttrue
\mciteSetBstMidEndSepPunct{\mcitedefaultmidpunct}
{\mcitedefaultendpunct}{\mcitedefaultseppunct}\relax
\EndOfBibitem
\bibitem[Erdahl(1979)]{Erdahl1979}
Erdahl,~R.~M. Two Algorithms for the Lower Bound Method of Reduced Density
  Matrix Theory. \emph{Rep. Math. Phys.} \textbf{1979}, \emph{15},
  147--162\relax
\mciteBstWouldAddEndPuncttrue
\mciteSetBstMidEndSepPunct{\mcitedefaultmidpunct}
{\mcitedefaultendpunct}{\mcitedefaultseppunct}\relax
\EndOfBibitem
\bibitem[Mazziotti and Erdahl(2001)Mazziotti, and Erdahl]{Mazziotti2001a}
Mazziotti,~D.~A.; Erdahl,~R.~M. Uncertainty Relations and Reduced Density
  Matrices: Mapping Many-Body Quantum Mechanics onto Four Particles.
  \emph{Phys. Rev. A} \textbf{2001}, \emph{63}, 042113\relax
\mciteBstWouldAddEndPuncttrue
\mciteSetBstMidEndSepPunct{\mcitedefaultmidpunct}
{\mcitedefaultendpunct}{\mcitedefaultseppunct}\relax
\EndOfBibitem
\bibitem[Nakata \latin{et~al.}(2001)Nakata, Nakatsuji, Ehara, Fukuda, Nakata,
  and Fujisawa]{Nakata2001}
Nakata,~M.; Nakatsuji,~H.; Ehara,~M.; Fukuda,~M.; Nakata,~K.; Fujisawa,~K.
  Variational Calculations of Fermion Second-Order Reduced Density Matrices by
  Semidefinite Programming Algorithm. \emph{J. Chem. Phys.} \textbf{2001},
  \emph{114}, 8282--8292\relax
\mciteBstWouldAddEndPuncttrue
\mciteSetBstMidEndSepPunct{\mcitedefaultmidpunct}
{\mcitedefaultendpunct}{\mcitedefaultseppunct}\relax
\EndOfBibitem
\bibitem[Zhao \latin{et~al.}(2004)Zhao, Braams, Fukuda, Overton, and
  Percus]{Zhao2004}
Zhao,~Z.; Braams,~B.; Fukuda,~M.; Overton,~M.; Percus,~J. The Reduced Density
  Matrix Method for Electronic Structure Calculations and the Role of
  Three-Index Representability Conditions. \emph{J. Chem. Phys.} \textbf{2004},
  \emph{120}, 2095--2104\relax
\mciteBstWouldAddEndPuncttrue
\mciteSetBstMidEndSepPunct{\mcitedefaultmidpunct}
{\mcitedefaultendpunct}{\mcitedefaultseppunct}\relax
\EndOfBibitem
\bibitem[Mazziotti(2002)]{Mazziotti2002}
Mazziotti,~D.~A. Variational Minimization of Atomic and Molecular Ground-State
  Energies via the Two-Particle Reduced Density Matrix. \emph{Phys. Rev. A}
  \textbf{2002}, \emph{65}, 062511\relax
\mciteBstWouldAddEndPuncttrue
\mciteSetBstMidEndSepPunct{\mcitedefaultmidpunct}
{\mcitedefaultendpunct}{\mcitedefaultseppunct}\relax
\EndOfBibitem
\bibitem[Mazziotti(2006)]{Mazziotti2006}
Mazziotti,~D.~A. Variational Reduced-Density-Matrix Method using Three-Particle
  $N$-Representability Conditions with Application to Many-Electron Molecules.
  \emph{Phys. Rev. A} \textbf{2006}, \emph{74}, 032501\relax
\mciteBstWouldAddEndPuncttrue
\mciteSetBstMidEndSepPunct{\mcitedefaultmidpunct}
{\mcitedefaultendpunct}{\mcitedefaultseppunct}\relax
\EndOfBibitem
\bibitem[Gidofalvi and Mazziotti(2008)Gidofalvi, and Mazziotti]{Gidofalvi2008}
Gidofalvi,~G.; Mazziotti,~D.~A. Active-Space Two-Electron
  Reduced-Density-Matrix Method: Complete Active-Space Calculations without
  Diagonalization of the $N$-Electron Hamiltonian. \emph{J. Chem. Phys.}
  \textbf{2008}, \emph{129}, 134108\relax
\mciteBstWouldAddEndPuncttrue
\mciteSetBstMidEndSepPunct{\mcitedefaultmidpunct}
{\mcitedefaultendpunct}{\mcitedefaultseppunct}\relax
\EndOfBibitem
\bibitem[Pelzer \latin{et~al.}(2011)Pelzer, Greenman, Gidofalvi, and
  Mazziotti]{Pelzer2011}
Pelzer,~K.; Greenman,~L.; Gidofalvi,~G.; Mazziotti,~D.~A. Strong Correlation in
  Acene Sheets from the Active-Space Variational Two-Electron Reduced Density
  Matrix Method: Effects of Symmetry and Size. \emph{J. Phys. Chem. A}
  \textbf{2011}, \emph{115}, 5632--5640\relax
\mciteBstWouldAddEndPuncttrue
\mciteSetBstMidEndSepPunct{\mcitedefaultmidpunct}
{\mcitedefaultendpunct}{\mcitedefaultseppunct}\relax
\EndOfBibitem
\bibitem[Verstichel \latin{et~al.}(2012)Verstichel, van Aggelen, Poelmans, and
  Van~Neck]{Verstichel2012}
Verstichel,~B.; van Aggelen,~H.; Poelmans,~W.; Van~Neck,~D. Variational
  Two-Particle Density Matrix Calculation for the Hubbard Model Below Half
  Filling Using Spin-Adapted Lifting Conditions. \emph{Phys. Rev. Lett.}
  \textbf{2012}, \emph{108}, 213001\relax
\mciteBstWouldAddEndPuncttrue
\mciteSetBstMidEndSepPunct{\mcitedefaultmidpunct}
{\mcitedefaultendpunct}{\mcitedefaultseppunct}\relax
\EndOfBibitem
\bibitem[Fosso-Tande \latin{et~al.}(2016)Fosso-Tande, Nguyen, Gidofalvi, and
  DePrince]{Fosso-Tande2016}
Fosso-Tande,~J.; Nguyen,~T.~S.; Gidofalvi,~G.; DePrince,~A.~E.,~III Large-Scale
  Variational Two-Electron Reduced-Density-Matrix-Driven Complete Active Space
  Self-Consistent Field Methods. \emph{J. Chem. Theory Comput.} \textbf{2016},
  \emph{12}, 2260--2271\relax
\mciteBstWouldAddEndPuncttrue
\mciteSetBstMidEndSepPunct{\mcitedefaultmidpunct}
{\mcitedefaultendpunct}{\mcitedefaultseppunct}\relax
\EndOfBibitem
\bibitem[Mazziotti(2016)]{Mazziotti2016}
Mazziotti,~D.~A. Enhanced Constraints for Accurate Lower Bounds on
  Many-Electron Quantum Energies from Variational Two-Electron Reduced Density
  Matrix Theory. \emph{Phys. Rev. Lett.} \textbf{2016}, \emph{117},
  153001\relax
\mciteBstWouldAddEndPuncttrue
\mciteSetBstMidEndSepPunct{\mcitedefaultmidpunct}
{\mcitedefaultendpunct}{\mcitedefaultseppunct}\relax
\EndOfBibitem
\bibitem[Coleman(1963)]{Coleman1963}
Coleman,~A.~J. Structure of Fermion Density Matrices. \emph{Rev. Mod. Phys.}
  \textbf{1963}, \emph{35}, 668\relax
\mciteBstWouldAddEndPuncttrue
\mciteSetBstMidEndSepPunct{\mcitedefaultmidpunct}
{\mcitedefaultendpunct}{\mcitedefaultseppunct}\relax
\EndOfBibitem
\bibitem[Erdahl(1978)]{Erdahl1978}
Erdahl,~R.~M. Representability. \emph{Int. J. Quantum Chem.} \textbf{1978},
  \emph{13}, 697--718\relax
\mciteBstWouldAddEndPuncttrue
\mciteSetBstMidEndSepPunct{\mcitedefaultmidpunct}
{\mcitedefaultendpunct}{\mcitedefaultseppunct}\relax
\EndOfBibitem
\bibitem[Kummer(1967)]{Kummer1967}
Kummer,~H. Eta-Representability Problem for Reduced Density Matrices. \emph{J.
  Math. Phys.} \textbf{1967}, \emph{8}, 2063\relax
\mciteBstWouldAddEndPuncttrue
\mciteSetBstMidEndSepPunct{\mcitedefaultmidpunct}
{\mcitedefaultendpunct}{\mcitedefaultseppunct}\relax
\EndOfBibitem
\bibitem[Vandenberghe and Boyd(1996)Vandenberghe, and Boyd]{Vandenberghe1996}
Vandenberghe,~L.; Boyd,~S. Semidefinite Programming. \emph{SIAM Rev.}
  \textbf{1996}, \emph{38}, 49--95\relax
\mciteBstWouldAddEndPuncttrue
\mciteSetBstMidEndSepPunct{\mcitedefaultmidpunct}
{\mcitedefaultendpunct}{\mcitedefaultseppunct}\relax
\EndOfBibitem
\bibitem[Mazziotti(2004)]{Mazziotti2004}
Mazziotti,~D.~A. Realization of Quantum Chemistry without Wave Functions
  through First-Order Semidefinite Programming. \emph{Phys. Rev. Lett.}
  \textbf{2004}, \emph{93}, 213001\relax
\mciteBstWouldAddEndPuncttrue
\mciteSetBstMidEndSepPunct{\mcitedefaultmidpunct}
{\mcitedefaultendpunct}{\mcitedefaultseppunct}\relax
\EndOfBibitem
\bibitem[Mazziotti(2007)]{Mazziotti2007}
Mazziotti,~D.~A. First-Order Semidefinite Programming for the Two-Electron
  Treatment of Many-Electron Atoms and Molecules. \emph{Math. Model. Numer.
  Anal.} \textbf{2007}, \emph{41}, 249--259\relax
\mciteBstWouldAddEndPuncttrue
\mciteSetBstMidEndSepPunct{\mcitedefaultmidpunct}
{\mcitedefaultendpunct}{\mcitedefaultseppunct}\relax
\EndOfBibitem
\bibitem[Mazziotti(2011)]{Mazziotti2011}
Mazziotti,~D.~A. Large-Scale Semidefinite Programming for Many-Electron Quantum
  Mechanics. \emph{Phys. Rev. Lett.} \textbf{2011}, \emph{106}, 083001\relax
\mciteBstWouldAddEndPuncttrue
\mciteSetBstMidEndSepPunct{\mcitedefaultmidpunct}
{\mcitedefaultendpunct}{\mcitedefaultseppunct}\relax
\EndOfBibitem
\bibitem[Piris(2013)]{Piris2013}
Piris,~M. Bounds on the PNOF5 Natural Geminal Occupation Numbers. \emph{Comput.
  and Theor. Chem.} \textbf{2013}, \emph{1003}, 123--126\relax
\mciteBstWouldAddEndPuncttrue
\mciteSetBstMidEndSepPunct{\mcitedefaultmidpunct}
{\mcitedefaultendpunct}{\mcitedefaultseppunct}\relax
\EndOfBibitem
\bibitem[Poelmans \latin{et~al.}(2015)Poelmans, Van~Raemdonck, Verstichel,
  De~Baerdemacker, Torre, Lain, Massaccesi, Alcoba, Bultinck, and
  Van~Neck]{Poelmans2015}
Poelmans,~W.; Van~Raemdonck,~M.; Verstichel,~B.; De~Baerdemacker,~S.;
  Torre,~A.; Lain,~L.; Massaccesi,~G.~E.; Alcoba,~D.~R.; Bultinck,~P.;
  Van~Neck,~D. Variational Optimization of the Second-Order Density Matrix
  Corresponding to a Seniority-Zero Configuration Interaction Wave Function.
  \emph{J. Chem. Theory Comput.} \textbf{2015}, \emph{11}, 4064--4076\relax
\mciteBstWouldAddEndPuncttrue
\mciteSetBstMidEndSepPunct{\mcitedefaultmidpunct}
{\mcitedefaultendpunct}{\mcitedefaultseppunct}\relax
\EndOfBibitem
\bibitem[Naftchi-Ardebili \latin{et~al.}(2011)Naftchi-Ardebili, Hau, and
  Mazziotti]{Naftchi-Ardebili2011}
Naftchi-Ardebili,~K.; Hau,~N.~W.; Mazziotti,~D.~A. Rank Restriction for the
  Variational Calculation of Two-Electron Reduced Density Matrices of
  Many-Electron Atoms and Molecules. \emph{Phys. Rev. A} \textbf{2011},
  \emph{84}, 052506\relax
\mciteBstWouldAddEndPuncttrue
\mciteSetBstMidEndSepPunct{\mcitedefaultmidpunct}
{\mcitedefaultendpunct}{\mcitedefaultseppunct}\relax
\EndOfBibitem
\bibitem[Head-Marsden and Mazziotti(2017)Head-Marsden, and
  Mazziotti]{Head-Marsden2017}
Head-Marsden,~K.; Mazziotti,~D.~A. Pair 2-Electron Reduced Density Matrix
  Theory Using Localized Orbitals. \emph{J. Chem. Phys.} \textbf{2017},
  \emph{147}, 084101\relax
\mciteBstWouldAddEndPuncttrue
\mciteSetBstMidEndSepPunct{\mcitedefaultmidpunct}
{\mcitedefaultendpunct}{\mcitedefaultseppunct}\relax
\EndOfBibitem
\bibitem[Alcoba \latin{et~al.}(2018)Alcoba, Torre, Lain, Massaccesi, O\~{n}a,
  Honor\'{e}, Poelmans, Van~Neck, Bultinck, and De~Baerdemacher]{Alcoba2018}
Alcoba,~D.~R.; Torre,~A.; Lain,~L.; Massaccesi,~G.~E.; O\~{n}a,~O.~B.;
  Honor\'{e},~E.~M.; Poelmans,~W.; Van~Neck,~D.; Bultinck,~P.;
  De~Baerdemacher,~S. Direct Variational Determination of the Two-Electron
  Reduced Density Matrix for Doubly Occupied-Configuration Interaction Wave
  Functions: The Influence of Three-Index $N$-Representability Conditions.
  \emph{J. Chem. Phys.} \textbf{2018}, \emph{148}, 024105\relax
\mciteBstWouldAddEndPuncttrue
\mciteSetBstMidEndSepPunct{\mcitedefaultmidpunct}
{\mcitedefaultendpunct}{\mcitedefaultseppunct}\relax
\EndOfBibitem
\bibitem[Alcoba \latin{et~al.}(2018)Alcoba, Capuzzi, Rubio-Garcia, Dukelsky,
  Massaccesi, O\~na, Torre, and Lain]{Alcoba2018_2}
Alcoba,~D.~R.; Capuzzi,~P.; Rubio-Garcia,~A.; Dukelsky,~J.; Massaccesi,~G.~E.;
  O\~na,~O.~B.; Torre,~A.; Lain,~L. Variational Reduced Density Matrix Method
  in the Doubly Occupied Configuration Interaction Space using Three-Particle
  $N$-Representability Conditions. \emph{J. Chem. Phys.} \textbf{2018},
  \emph{149}, 194105\relax
\mciteBstWouldAddEndPuncttrue
\mciteSetBstMidEndSepPunct{\mcitedefaultmidpunct}
{\mcitedefaultendpunct}{\mcitedefaultseppunct}\relax
\EndOfBibitem
\bibitem[Alcoba \latin{et~al.}(2019)Alcoba, Torre, Lain, Massaccesi, O\~na, and
  R\`{\i}os]{Alcoba2019}
Alcoba,~D.~R.; Torre,~A.; Lain,~L.; Massaccesi,~G.~E.; O\~na,~O.~B.;
  R\`{\i}os,~E. Unrestricted Treatment for the Direct Variational Determination
  of the Two-Electron Reduced Density Matrix for Doubly
  Occupied-Configuration-Interaction Wave Functions. \emph{J. Chem. Phys.}
  \textbf{2019}, \emph{150}, 164106\relax
\mciteBstWouldAddEndPuncttrue
\mciteSetBstMidEndSepPunct{\mcitedefaultmidpunct}
{\mcitedefaultendpunct}{\mcitedefaultseppunct}\relax
\EndOfBibitem
\bibitem[Weinhold and {Wilson Jr.}(1967)Weinhold, and {Wilson
  Jr.}]{Weinhold1967}
Weinhold,~F.; {Wilson Jr.},~E.~B. Reduced Density Matrics of Atoms and
  Molecules. I. The 2 Matrix of Double-Occupancy, Configuration-Interaction
  Wavefunctions for Singlet States. \emph{J. Chem. Phys.} \textbf{1967},
  \emph{46}, 2752\relax
\mciteBstWouldAddEndPuncttrue
\mciteSetBstMidEndSepPunct{\mcitedefaultmidpunct}
{\mcitedefaultendpunct}{\mcitedefaultseppunct}\relax
\EndOfBibitem
\bibitem[Szabo and Ostlund(1996)Szabo, and Ostlund]{Szabo1996}
Szabo,~A.; Ostlund,~N.~S. \emph{Modern Quantum Chemistry: Introduction to
  Advanced Electronic Structure Theory}; Dover: New York, 1996\relax
\mciteBstWouldAddEndPuncttrue
\mciteSetBstMidEndSepPunct{\mcitedefaultmidpunct}
{\mcitedefaultendpunct}{\mcitedefaultseppunct}\relax
\EndOfBibitem
\bibitem[Limacher \latin{et~al.}(2013)Limacher, Ayers, Johnson,
  De~Baerdemacker, Van~Neck, and Bultinck]{Limacher2013}
Limacher,~P.~A.; Ayers,~P.~W.; Johnson,~P.~A.; De~Baerdemacker,~S.;
  Van~Neck,~D.; Bultinck,~P. A New Mean-Field Method Suitable for Strongly
  Correlated Electrons: Computationally Facile Antisymmetric Products of
  Nonorthogonal Geminals. \emph{J. Chem. Theory Comput.} \textbf{2013},
  \emph{9}, 1394--1401\relax
\mciteBstWouldAddEndPuncttrue
\mciteSetBstMidEndSepPunct{\mcitedefaultmidpunct}
{\mcitedefaultendpunct}{\mcitedefaultseppunct}\relax
\EndOfBibitem
\bibitem[Boguslawski \latin{et~al.}(2014)Boguslawski, Tecmer, Limacher,
  Johnson, Ayers, Bultinck, Baerdemacker, and {Van Neck}]{Boguslawski2014}
Boguslawski,~K.; Tecmer,~P.; Limacher,~P.~A.; Johnson,~P.~A.; Ayers,~P.~W.;
  Bultinck,~P.; Baerdemacker,~S.~D.; {Van Neck},~D. Projected Seniority-Two
  Orbital Optimization of the Antisymmetric Product of One-Reference Orbital
  Geminal. \emph{J. Chem. Phys.} \textbf{2014}, \emph{140}, 214114\relax
\mciteBstWouldAddEndPuncttrue
\mciteSetBstMidEndSepPunct{\mcitedefaultmidpunct}
{\mcitedefaultendpunct}{\mcitedefaultseppunct}\relax
\EndOfBibitem
\bibitem[Boguslawski \latin{et~al.}(2014)Boguslawski, Tecmer, Ayers, Bultinck,
  De~Baerdemacker, and {Van Neck}]{Boguslawski2014a}
Boguslawski,~K.; Tecmer,~P.; Ayers,~P.~W.; Bultinck,~P.; De~Baerdemacker,~S.;
  {Van Neck},~D. Efficient Description of Strongly Correlated Electrons with
  Mean-Field Cost. \emph{Phys. Rev. B} \textbf{2014}, \emph{89}, 201106\relax
\mciteBstWouldAddEndPuncttrue
\mciteSetBstMidEndSepPunct{\mcitedefaultmidpunct}
{\mcitedefaultendpunct}{\mcitedefaultseppunct}\relax
\EndOfBibitem
\bibitem[Tecmer \latin{et~al.}(2014)Tecmer, Boguslawski, Johnson, Limacher,
  Chan, Verstraelen, and Ayers]{Tecmer2014}
Tecmer,~P.; Boguslawski,~K.; Johnson,~P.~A.; Limacher,~P.~A.; Chan,~M.;
  Verstraelen,~T.; Ayers,~P.~W. Assessing the Accuracy of New Geminal-Based
  Approaches. \emph{J. Phys. Chem. A} \textbf{2014}, \emph{118},
  9058--9068\relax
\mciteBstWouldAddEndPuncttrue
\mciteSetBstMidEndSepPunct{\mcitedefaultmidpunct}
{\mcitedefaultendpunct}{\mcitedefaultseppunct}\relax
\EndOfBibitem
\bibitem[Boguslawski and Ayers(2015)Boguslawski, and Ayers]{Boguslawski2015}
Boguslawski,~K.; Ayers,~P.~W. Linearized Coupled Cluster Correction on the
  Antisymmetric Product of 1-Reference Orbital Geminals. \emph{J. Chem. Theory
  Comput.} \textbf{2015}, \emph{11}, 5252--5261\relax
\mciteBstWouldAddEndPuncttrue
\mciteSetBstMidEndSepPunct{\mcitedefaultmidpunct}
{\mcitedefaultendpunct}{\mcitedefaultseppunct}\relax
\EndOfBibitem
\bibitem[Bytautas \latin{et~al.}(2011)Bytautas, Henderson, Jiménez-Hoyos,
  Ellis, and Scuseria]{Bytautas2011}
Bytautas,~L.; Henderson,~T.~M.; Jiménez-Hoyos,~C.~A.; Ellis,~J.~K.;
  Scuseria,~G.~E. Seniority and Orbital Symmetry as Tools for Establishing a
  Full Configuration Interaction Hierarchy. \emph{J. Chem. Phys.}
  \textbf{2011}, \emph{135}, 044119\relax
\mciteBstWouldAddEndPuncttrue
\mciteSetBstMidEndSepPunct{\mcitedefaultmidpunct}
{\mcitedefaultendpunct}{\mcitedefaultseppunct}\relax
\EndOfBibitem
\bibitem[Stein \latin{et~al.}(2014)Stein, Henderson, and Scuseria]{Stein2014}
Stein,~T.; Henderson,~T.~M.; Scuseria,~G.~E. Seniority Zero Pair Coupled
  Cluster Doubles Theory. \emph{J. Chem. Phys.} \textbf{2014}, \emph{140},
  214113\relax
\mciteBstWouldAddEndPuncttrue
\mciteSetBstMidEndSepPunct{\mcitedefaultmidpunct}
{\mcitedefaultendpunct}{\mcitedefaultseppunct}\relax
\EndOfBibitem
\bibitem[Henderson \latin{et~al.}(2014)Henderson, Bulik, Stein, and
  Scuseria]{Henderson2014}
Henderson,~T.~M.; Bulik,~I.~W.; Stein,~T.; Scuseria,~G.~E. Seniority-Based
  Coupled Cluster Theory. \emph{J. Chem. Phys.} \textbf{2014}, \emph{141},
  244104\relax
\mciteBstWouldAddEndPuncttrue
\mciteSetBstMidEndSepPunct{\mcitedefaultmidpunct}
{\mcitedefaultendpunct}{\mcitedefaultseppunct}\relax
\EndOfBibitem
\bibitem[Henderson \latin{et~al.}(2015)Henderson, Bulik, and
  Scuseria]{Henderson2015}
Henderson,~T.~M.; Bulik,~I.~W.; Scuseria,~G.~E. Pair Extended Coupled Cluster
  Doubles. \emph{J. Chem. Phys.} \textbf{2015}, \emph{142}, 214116\relax
\mciteBstWouldAddEndPuncttrue
\mciteSetBstMidEndSepPunct{\mcitedefaultmidpunct}
{\mcitedefaultendpunct}{\mcitedefaultseppunct}\relax
\EndOfBibitem
\bibitem[Bulik \latin{et~al.}(2015)Bulik, Henderson, and Scuseria]{Bulik2015}
Bulik,~I.~W.; Henderson,~T.~M.; Scuseria,~G.~E. Can Single-Reference Coupled
  Cluster Theory Describe Static Correlation? \emph{J. Chem. Theory Comput.}
  \textbf{2015}, \emph{11}, 7, 3171--3179\relax
\mciteBstWouldAddEndPuncttrue
\mciteSetBstMidEndSepPunct{\mcitedefaultmidpunct}
{\mcitedefaultendpunct}{\mcitedefaultseppunct}\relax
\EndOfBibitem
\bibitem[Shepherd \latin{et~al.}(2016)Shepherd, Henderson, and
  Scuseria]{Shepherd2016}
Shepherd,~J.~J.; Henderson,~T.~M.; Scuseria,~G.~E. Using Full Configuration
  Interaction Quantum Monte Carlo in a Seniority Zero Space to Investigate the
  Correlation Energy Equivalence of Pair Coupled Cluster Doubles and Doubly
  Occupied Configuration Interaction. \emph{J. Chem. Phys.} \textbf{2016},
  \emph{144}, 094112\relax
\mciteBstWouldAddEndPuncttrue
\mciteSetBstMidEndSepPunct{\mcitedefaultmidpunct}
{\mcitedefaultendpunct}{\mcitedefaultseppunct}\relax
\EndOfBibitem
\bibitem[Bytautas \latin{et~al.}(2011)Bytautas, Henderson, Jim\'{e}nez-Hoyos,
  Ellis, and Scuseria]{Byautas2011}
Bytautas,~L.; Henderson,~T.~M.; Jim\'{e}nez-Hoyos,~C.~A.; Ellis,~J.~K.;
  Scuseria,~G.~E. Seniority and Orbital Symmetry as Tools for Establishing a
  Full Configuration Interaction Hierarchy. \emph{J. Chem. Phys.}
  \textbf{2011}, \emph{135}, 044119\relax
\mciteBstWouldAddEndPuncttrue
\mciteSetBstMidEndSepPunct{\mcitedefaultmidpunct}
{\mcitedefaultendpunct}{\mcitedefaultseppunct}\relax
\EndOfBibitem
\bibitem[Mazziotti(2012)]{Mazziotti2012}
Mazziotti,~D.~A. Two-Electron Reduced Density Matrix as the Basic Variable in
  Many-Electron Quantum Chemistry and Physics. \emph{Chem. Rev.} \textbf{2012},
  \emph{112}, 244--262\relax
\mciteBstWouldAddEndPuncttrue
\mciteSetBstMidEndSepPunct{\mcitedefaultmidpunct}
{\mcitedefaultendpunct}{\mcitedefaultseppunct}\relax
\EndOfBibitem
\bibitem[Coleman and Yukalov(2000)Coleman, and Yukalov]{Coleman2000}
Coleman,~A.~J.; Yukalov,~V.~I. \emph{Reduced Density Matrices: Coulson's
  Challenge, Chapter 4}; Springer, 2000\relax
\mciteBstWouldAddEndPuncttrue
\mciteSetBstMidEndSepPunct{\mcitedefaultmidpunct}
{\mcitedefaultendpunct}{\mcitedefaultseppunct}\relax
\EndOfBibitem
\bibitem[Davidson(1976)]{Davidson1976}
Davidson,~E.~R. \emph{Reduced Density Matrices in Quantum Chemistry}; Academic:
  New York, 1976\relax
\mciteBstWouldAddEndPuncttrue
\mciteSetBstMidEndSepPunct{\mcitedefaultmidpunct}
{\mcitedefaultendpunct}{\mcitedefaultseppunct}\relax
\EndOfBibitem
\bibitem[Valdemoro(1992)]{Valdemoro1992}
Valdemoro,~C. Approximation the 2nd-Order Reduced Density-Matrix in Terms of
  the 1st-Order One. \emph{Phys. Rev. A} \textbf{1992}, \emph{45},
  4462--4467\relax
\mciteBstWouldAddEndPuncttrue
\mciteSetBstMidEndSepPunct{\mcitedefaultmidpunct}
{\mcitedefaultendpunct}{\mcitedefaultseppunct}\relax
\EndOfBibitem
\bibitem[Nakatsuji and Yasuda(1996)Nakatsuji, and Yasuda]{Nakatsuji1996}
Nakatsuji,~H.; Yasuda,~K. Direct Determination of the Quantum-Mechanical
  Density Matrix using the Density Equation. \emph{Phys. Rev. Lett.}
  \textbf{1996}, \emph{76}, 1039--1042\relax
\mciteBstWouldAddEndPuncttrue
\mciteSetBstMidEndSepPunct{\mcitedefaultmidpunct}
{\mcitedefaultendpunct}{\mcitedefaultseppunct}\relax
\EndOfBibitem
\bibitem[Mazziotti(1998)]{Mazziotti1998}
Mazziotti,~D.~A. Contracted Schrodinger Equation: Determining Quantum Energies
  and Two-Particle Density Matrices without Wave Functions. \emph{Phys. Rev. A}
  \textbf{1998}, \emph{57}, 4219--4234\relax
\mciteBstWouldAddEndPuncttrue
\mciteSetBstMidEndSepPunct{\mcitedefaultmidpunct}
{\mcitedefaultendpunct}{\mcitedefaultseppunct}\relax
\EndOfBibitem
\bibitem[L{\"o}wdin(1955)]{Lowdin1955}
L{\"o}wdin,~P.~O. Quantum Theory of Many-Particle Systems .1. Physical
  Interpretations by Means of Density Matrices, Natural Spin-Orbitals, and
  Convergence Problems in the Method of Configuration Interaction. \emph{Phys.
  Rev.} \textbf{1955}, \emph{97}, 1474--1489\relax
\mciteBstWouldAddEndPuncttrue
\mciteSetBstMidEndSepPunct{\mcitedefaultmidpunct}
{\mcitedefaultendpunct}{\mcitedefaultseppunct}\relax
\EndOfBibitem
\bibitem[Mayer(1955)]{Mayer1955}
Mayer,~J.~E. Electron Correlation. \emph{Phys. Rev.} \textbf{1955}, \emph{100},
  1579--1586\relax
\mciteBstWouldAddEndPuncttrue
\mciteSetBstMidEndSepPunct{\mcitedefaultmidpunct}
{\mcitedefaultendpunct}{\mcitedefaultseppunct}\relax
\EndOfBibitem
\bibitem[Slebodziński(1970)]{Slebodzinski1970}
Slebodziński,~W. \emph{Exterior Forms and their Applications}; Polish
  Scientific Publishers: Warsaw, 1970\relax
\mciteBstWouldAddEndPuncttrue
\mciteSetBstMidEndSepPunct{\mcitedefaultmidpunct}
{\mcitedefaultendpunct}{\mcitedefaultseppunct}\relax
\EndOfBibitem
\bibitem[Mazziotti(2012)]{Mazziotti2012a}
Mazziotti,~D.~A. Structure of Fermionic Density Matrices: Complete
  N-Representability Conditions. \emph{Phys. Rev. Lett.} \textbf{2012},
  \emph{108}, 263002\relax
\mciteBstWouldAddEndPuncttrue
\mciteSetBstMidEndSepPunct{\mcitedefaultmidpunct}
{\mcitedefaultendpunct}{\mcitedefaultseppunct}\relax
\EndOfBibitem
\bibitem[Garrod and Percus(1964)Garrod, and Percus]{Garrod1964}
Garrod,~C.; Percus,~J. Reduction of $N$-Particle Variational Problem. \emph{J.
  Math. Phys.} \textbf{1964}, \emph{5}, 1756\relax
\mciteBstWouldAddEndPuncttrue
\mciteSetBstMidEndSepPunct{\mcitedefaultmidpunct}
{\mcitedefaultendpunct}{\mcitedefaultseppunct}\relax
\EndOfBibitem
\bibitem[Fukuda \latin{et~al.}(2007)Fukuda, Braams, Nakata, Overton, Percus,
  Yamashita, and Zhao]{Fukuda2007}
Fukuda,~M.; Braams,~B.~J.; Nakata,~M.; Overton,~M.~L.; Percus,~J.~K.;
  Yamashita,~M.; Zhao,~Z. Large-Scale Semidefinite Programs in Electronic
  Structure Calculation. \emph{Math. Program.} \textbf{2007}, \emph{109},
  553--580\relax
\mciteBstWouldAddEndPuncttrue
\mciteSetBstMidEndSepPunct{\mcitedefaultmidpunct}
{\mcitedefaultendpunct}{\mcitedefaultseppunct}\relax
\EndOfBibitem
\bibitem[Roos(1987)]{Roos1987}
Roos,~B.~O. In \emph{Ab Initio Methods in Quantum Chemistry II}; Lawly,~K.~P.,
  Ed.; Wiley: New York, 1987; pp 399--446\relax
\mciteBstWouldAddEndPuncttrue
\mciteSetBstMidEndSepPunct{\mcitedefaultmidpunct}
{\mcitedefaultendpunct}{\mcitedefaultseppunct}\relax
\EndOfBibitem
\bibitem[Werner and Knowles(1985)Werner, and Knowles]{Werner1985}
Werner,~H.-J.; Knowles,~P.~J. A Second Order Multiconfiguration
  Self-Consistent-Field Procedure with Optimum Convergence. \emph{J. Chem.
  Phys.} \textbf{1985}, \emph{82 (11)}, 5053--5063\relax
\mciteBstWouldAddEndPuncttrue
\mciteSetBstMidEndSepPunct{\mcitedefaultmidpunct}
{\mcitedefaultendpunct}{\mcitedefaultseppunct}\relax
\EndOfBibitem
\bibitem[Sun \latin{et~al.}(2017)Sun, Yang, and Chan]{Sun2017}
Sun,~Q.; Yang,~J.; Chan,~G.~K. A General Second Order Complete Active Space
  Self-Consistent-Field Solver for Large-Scale Systems. \emph{Chemical Physics
  Letters} \textbf{2017}, \emph{683}, 291--299\relax
\mciteBstWouldAddEndPuncttrue
\mciteSetBstMidEndSepPunct{\mcitedefaultmidpunct}
{\mcitedefaultendpunct}{\mcitedefaultseppunct}\relax
\EndOfBibitem
\bibitem[RDM(2019)]{RDMChem}
\emph{Maple Quantum Chemistry Toolbox}; RDMChem, Chicago, 2019\relax
\mciteBstWouldAddEndPuncttrue
\mciteSetBstMidEndSepPunct{\mcitedefaultmidpunct}
{\mcitedefaultendpunct}{\mcitedefaultseppunct}\relax
\EndOfBibitem
\bibitem[Map(2019)]{Maple2019}
\emph{Maple 2019}; Maplesoft, Waterloo, 2019\relax
\mciteBstWouldAddEndPuncttrue
\mciteSetBstMidEndSepPunct{\mcitedefaultmidpunct}
{\mcitedefaultendpunct}{\mcitedefaultseppunct}\relax
\EndOfBibitem
\bibitem[{Dunning Jr.}(1989)]{Dunning1989}
{Dunning Jr.},~T.~H. Gaussian-Basis Sets for Use in Correlated Molecular
  Calculations .1. The Atoms Boron through Neon and Hydrogen. \emph{J. Chem.
  Phys.} \textbf{1989}, \emph{90}, 1007--1023\relax
\mciteBstWouldAddEndPuncttrue
\mciteSetBstMidEndSepPunct{\mcitedefaultmidpunct}
{\mcitedefaultendpunct}{\mcitedefaultseppunct}\relax
\EndOfBibitem
\bibitem[Hehre \latin{et~al.}(1972)Hehre, Ditchfield, and Pople]{Hehre1972}
Hehre,~W.~J.; Ditchfield,~R.; Pople,~J.~A. Self-Consistent Molecular-Orbital
  Methods .12. Further Extensions of Gaussian-Type Basis Sets for Use in
  Molecular-Orbital Studies of Organic-Molecules. \emph{J. Chem. Phys.}
  \textbf{1972}, \emph{56}, 2257\relax
\mciteBstWouldAddEndPuncttrue
\mciteSetBstMidEndSepPunct{\mcitedefaultmidpunct}
{\mcitedefaultendpunct}{\mcitedefaultseppunct}\relax
\EndOfBibitem
\bibitem[Hehre \latin{et~al.}(1969)Hehre, Stewart, and Pople]{Hehre1969}
Hehre,~W.~J.; Stewart,~R.~F.; Pople,~J.~A. Self-Consistent Molecular-Orbital
  Methods .I. Use of Gaussian Expansions of Slate-Type Atomic Orbitals.
  \emph{J. Chem. Phys.} \textbf{1969}, \emph{51}, 2657\relax
\mciteBstWouldAddEndPuncttrue
\mciteSetBstMidEndSepPunct{\mcitedefaultmidpunct}
{\mcitedefaultendpunct}{\mcitedefaultseppunct}\relax
\EndOfBibitem
\bibitem[Pietro and Hehre(1983)Pietro, and Hehre]{Pietro1983}
Pietro,~W.~J.; Hehre,~W.~J. Molecular Orbital Theory of the Properties of
  Inorganic and Organometallic Compounds. 3. STO-3G Basis Sets for First- and
  Second-Row Transition Metals. \emph{Journal of Computational Chemistry}
  \textbf{1983}, \emph{4}, 241--251\relax
\mciteBstWouldAddEndPuncttrue
\mciteSetBstMidEndSepPunct{\mcitedefaultmidpunct}
{\mcitedefaultendpunct}{\mcitedefaultseppunct}\relax
\EndOfBibitem
\bibitem[Dobbs and Hehre(1987)Dobbs, and Hehre]{Dobbs1987}
Dobbs,~K.~D.; Hehre,~W.~J. Molecular-Orbital Theory of the Properties of
  Inorganic and Organometallic Compounds .6. Extended Basis-Sets for 2nd-Row
  Transition-Metals. \emph{J. Comput. Chem.} \textbf{1987}, \emph{8},
  880--893\relax
\mciteBstWouldAddEndPuncttrue
\mciteSetBstMidEndSepPunct{\mcitedefaultmidpunct}
{\mcitedefaultendpunct}{\mcitedefaultseppunct}\relax
\EndOfBibitem
\bibitem[Wenthold \latin{et~al.}(1998)Wenthold, Squires, and
  Lineberger]{Wenthold1998}
Wenthold,~P.~G.; Squires,~R.~R.; Lineberger,~W.~C. Ultraviolet Photoelectron
  Spectroscopy of the o-, m-, and p-Benzyne Negative Ions. Electron Affinities
  and Singlet−Triplet Splittings for o-, m-, and p-Benzyne. \emph{Journal of
  the American Chemical Society} \textbf{1998}, \emph{120}, 5279--5290\relax
\mciteBstWouldAddEndPuncttrue
\mciteSetBstMidEndSepPunct{\mcitedefaultmidpunct}
{\mcitedefaultendpunct}{\mcitedefaultseppunct}\relax
\EndOfBibitem
\bibitem[Shee \latin{et~al.}(2019)Shee, Arthur, Zhang, Reichman, and
  Friesner]{Shee2019}
Shee,~J.; Arthur,~E.~J.; Zhang,~S.; Reichman,~D.~R.; Friesner,~R.~A.
  Singlet–Triplet Energy Gaps of Organic Biradicals and Polyacenes with
  Auxiliary-Field Quantum Monte Carlo. \emph{Journal of Chemical Theory and
  Computation} \textbf{2019}, \emph{15}, 4924--4932, PMID: 31381324\relax
\mciteBstWouldAddEndPuncttrue
\mciteSetBstMidEndSepPunct{\mcitedefaultmidpunct}
{\mcitedefaultendpunct}{\mcitedefaultseppunct}\relax
\EndOfBibitem
\bibitem[Xie \latin{et~al.}(2020)Xie, Boyn, Filatov, McNeece, Mazziotti, and
  Anderson]{Xie2019}
Xie,~J.; Boyn,~J.-N.; Filatov,~A.~S.; McNeece,~A.~J.; Mazziotti,~D.~A.;
  Anderson,~J.~S. Redox, Transmetalation, and Stacking Properties of
  Tetrathiafulvalene-2,3,6,7-tetrathiolate Bridged Tin, Nickel, and Palladium
  Compounds. \emph{Chem. Sci.} \textbf{2020}, \emph{11}, 1066--1078\relax
\mciteBstWouldAddEndPuncttrue
\mciteSetBstMidEndSepPunct{\mcitedefaultmidpunct}
{\mcitedefaultendpunct}{\mcitedefaultseppunct}\relax
\EndOfBibitem
\bibitem[Burgess and Lowe(1996)Burgess, and Lowe]{Burgess1996}
Burgess,~B.~K.; Lowe,~D.~J. Mechanism of Molybdenum Nitrogenase. \emph{Chem.
  Rev.} \textbf{1996}, \emph{96}, 2983--2012\relax
\mciteBstWouldAddEndPuncttrue
\mciteSetBstMidEndSepPunct{\mcitedefaultmidpunct}
{\mcitedefaultendpunct}{\mcitedefaultseppunct}\relax
\EndOfBibitem
\bibitem[Hoffman \latin{et~al.}(2014)Hoffman, Lukoyanov, Yang, Dean, and
  Seefeldt]{Hoffman2014}
Hoffman,~B.~M.; Lukoyanov,~D.; Yang,~Z.-Y.; Dean,~D.~R.; Seefeldt,~L.~C.
  Mechanism of Nitrogen Fixation by Nitrogenase: The Next Stage. \emph{Chem.
  Rev.} \textbf{2014}, \emph{114}, 40414062\relax
\mciteBstWouldAddEndPuncttrue
\mciteSetBstMidEndSepPunct{\mcitedefaultmidpunct}
{\mcitedefaultendpunct}{\mcitedefaultseppunct}\relax
\EndOfBibitem
\end{mcitethebibliography}

\end{document}